\documentclass[conference]{IEEEtran}
\IEEEoverridecommandlockouts

\usepackage{cite}
\usepackage{amsmath,amssymb,amsfonts}
\usepackage{algorithmic}
\usepackage{graphicx}
\usepackage{subcaption}
\usepackage{textcomp}
\usepackage{xcolor}
\usepackage{booktabs}
\usepackage{orcidlink}
\newcommand{\namestrut}{\rule{0pt}{2.6ex}}
\usepackage{braket}
\usepackage{multirow}

\begin{document}

\title{Magic-Informed Quantum Architecture Search
}

\author{
\IEEEauthorblockN{
\namestrut Vincenzo Lipardi\,\orcidlink{0009-0000-3243-7969}
}
\IEEEauthorblockA{
\textit{Department of Advanced Computing Sciences} \\
\textit{Maastricht University} \\
Maastricht, The Netherlands \\
vincenzo.lipardi@maastrichtuniversity.nl
}
\and
\IEEEauthorblockN{
\namestrut Domenica Dibenedetto\,\orcidlink{0000-0002-2538-3170}
}
\IEEEauthorblockA{
\textit{Department of Advanced Computing Sciences} \\
\textit{Maastricht University} \\
 Maastricht, The Netherlands \\
\phantom{email@email.com}
}
\and
\IEEEauthorblockN{
\namestrut Georgios Stamoulis\,\orcidlink{0000-0001-7248-8197}
}
\IEEEauthorblockA{
\textit{Department of Advanced Computing Sciences} \\
\textit{Maastricht University} \\
Maastricht, The Netherlands \\
\phantom{email@email.com}
}
\and
\IEEEauthorblockN{
\namestrut Mark H. M. Winands\,\orcidlink{0000-0002-0125-0824}
}
\IEEEauthorblockA{
\textit{Department of Advanced Computing Sciences} \\
\textit{Maastricht University} \\
Maastricht, The Netherlands \\
\phantom{email@email.com}
}
}

\maketitle

\begin{abstract}

Nonstabilizerness, commonly referred to as magic, is a fundamental resource underpinning quantum advantage. 
In this paper, we propose a magic-informed quantum architecture search (QAS) technique that enables control over a quantum resource within the general framework of circuit design. Inspired by the AlphaGo approach, we tackle the problem with a Monte Carlo Tree Search technique equipped with a Graph Neural Network (GNN) that estimates the magic of candidate quantum circuits. The GNN model induces a magic-based bias that steers the search toward either high- or low-magic regimes, depending on the target objective.

We benchmark the proposed magic-informed QAS technique on both the structured ground-state energy problem and on the more general quantum state approximation problem, spanning different sizes and target magic levels.  
Experimental results show that the proposed technique effectively influences the magic across the search tree and notably also on the resulting final circuit, even in regimes where the GNN operates on out-of-distribution instances. Although introducing a problem-agnostic magic bias could, in principle, constrain the search dynamics, we observe consistent improvements in solution quality across all problems tested.

\end{abstract}

\begin{IEEEkeywords}
Quantum Architecture Search, Stabilizer Rényi Entropy, Monte Carlo Tree Search, Graph Neural Networks, Quantum Circuit Synthesis, Nonstabilizerness Estimation.
\end{IEEEkeywords}

\section{Introduction}
Quantum architecture search (QAS), also known as \textit{quantum ansatz search} or \textit{quantum circuit design}, is a general framework aiming to the development of techniques that automatically design parameterized quantum circuits (PQCs) tailored to both a target problem and the constraints of available quantum hardware. Despite the progress provided by several QAS techniques proposed~\cite{zhang2022differentiable, zhang2023evolutionary, chivilikhin2020mog, lourens2023hierarchical, kuo2021quantum, ostaszewski2021reinforcement, patel2024curriculum, wang, lipardi2025quantum, he2022quantum, he2025self}, designing PQCs that simultaneously achieve a high solution quality and satisfy practical hardware constraints remains a key challenge. This difficulty derives from the combinatorial challenge of selecting appropriate gates, positions, and parameters, as well as from the geometry of the optimization landscape, which is typically affected by the barren plateau~\cite{larocca2025barren,cerezo2021variational} or, more generally, the exponential concentration phenomenon~\cite{thanasilp2024exponential}.
In this scenario, exploiting problem-specific structure can significantly reduce the computational cost of QAS and improve the solution quality. For instance, in quantum chemistry applications, physics-inspired ansatz can incorporate molecular symmetries into the circuit design for solving the ground-state energy problem~\cite{grimsley2019adaptive}. However, in general applications, there is no trivial structural information to exploit and we have to rely on problem-agnostic QAS techniques. An example is the the Progressive Widening Monte Carlo Tree Search (PWMCTS)~\cite{lipardi2025quantum}, a gradient-free technique that exploits a progressive widening~\cite{chaslot} to dynamically explore the continuous search space of PQCs.

Beyond the combinatorial complexity of QAS, the design of PQCs also involves exploiting quantum resources. On the one hand, entanglement captures non-classical correlations, but it is not sufficient alone to guarantee computational advantage~\cite{gottesman1998heisenberg}. On the other hand nonstabilizerness, commonly referred to as magic, measures the deviation of a quantum state from the polynomial-time simulability in the framework of the Gottesman-Knill Theorem~\cite{gottesman1998heisenberg}. While several works take into account entanglement, we highlight the practical relevance of having access to magic estimations in the design of PQCs. 

In this paper, we propose the class of magic-informed QAS techniques aiming to add a problem-agnostic bias in the PQC design based on a quantum resource. 
The major obstacle to the practical use of magic in QAS techniques lies in the computational cost related to magic measures, referred to as magic monotones, as it generally grows exponentially in the number of qubits~\cite{leone2022stabilizer}. To overcome this limitation, we leverage a recently proposed Graph Neural Network (GNN) model~\cite{lipardi2026nonstabilizerness} to estimate the magic of PQCs, measured by the stabilizer $2$-Rényi entropy $M_2$~\cite{leone2022stabilizer}. The GNN model exhibits strong generalization capabilities across different PQCs in terms of number of qubits, circuit depth, circuit structures, and entanglement levels. Inspired by the AlphaGo approach~\cite{silver2016mastering}, which enhances Monte Carlo Tree Search with deep learning, we formulate the QAS problem as a single-player game and employ PWMCTS~\cite{lipardi2025quantum} alongside a GNN~\cite{lipardi2026nonstabilizerness} to predict the $M_2$ of candidate PQCs, inducing a bias toward desired magic regimes. 

The main contributions of this paper are as follows.
\begin{enumerate}
    \item We propose a magic-informed QAS technique to control the magic level during the design of PQCs. Depending on the scope, it can be used to bias the search toward either high-magic or low-magic circuits. In particular, we embed GNN-based $M_2$ estimations in the PWMCTS technique~\cite{lipardi2025quantum} at two different stages, including the expansion and selection strategy.
    
    \item We benchmark the magic-informed PWMCTS on both the structured ground-state energy problem and the more general quantum state approximation problem spanning different levels of magic, measured in terms of $M_2$. The proposed techniques effectively induce a bias on the $M_2$ values within the entire search tree and on the final PQC designed. Additionally, the solution quality achieved by the magic-informed PWMCTS is systematically improved.

    \item We provide a pipeline for designing target quantum circuits with different levels of magic under the constraint of a fixed number of gates. Keeping the total number of gates is fundamental to focus the hardness of the analysis on the magic levels rather than other PQC structures. We observe that the performance of standard PWMCTS significantly decreases as the magic level of the target circuits increases. In this regime, which is the most relevant for quantum advantage, magic-informed QAS techniques can benefit the design of more representative PQCs.
    
\end{enumerate}

It is important to note that quantum magic, as a quantum resource, naturally fits within the framework of problem-agnostic QAS techniques. Moreover, magic is attracting growing interest in many-body quantum physics~\cite{liu2022many, tarabunga2023many} and has been recently investigated in the context of variational quantum algorithms~\cite{spriggs2025quantum,capecci2025role}. In this scenario, magic-informed QAS techniques can help better study and exploit magic in quantum computing applications.

This paper is organized as follows. Section~\ref{quantum_magic} introduces quantum magic, the stabilizer Rényi entropy, and the GNN model employed for magic estimation. Section~\ref{qas} presents PWMCTS, focusing on its suitability for incorporating magic-based bias. Section~\ref{magic_informed_pwmcts} presents the details of the proposed magic-informed PWMCTS technique. Section~\ref{experiments} describes the experimental setup and results. Section~\ref{discussion} discusses potentials and limitations of magic-informed QAS. Finally, Section~\ref{conclusions} summarizes the main findings and outlines future research.

\section{Quantum Magic}\label{quantum_magic}
The Gottesman-Knill Theorem~\cite{gottesman1998heisenberg, aaronson2004improved} states that quantum states that can be prepared using exclusively Clifford gates, referred to as \textit{stabilizer states}, can be perfectly simulated in polynomial time on a classical computer. As a consequence, quantum computations restricted to stabilizer states and Clifford gates do not provide any computational advantage over the classical counterpart. Because the set of stabilizer states is closed under Clifford operations, and a universal quantum gate set cannot be constructed from Clifford gates alone, the preparation of a generic state in the Hilbert space requires resources beyond the Clifford group. The amount of these non-Clifford resources is referred to as \textit{magic}, or nonstabilizerness. 

In recent years, magic has been extensively studied. On the one hand, considerable effort has been dedicated to define proper magic monotones~\cite{heinrich2019robustness, leone2022stabilizer, haug2023stabilizer, sonya2025nonstabilizerness, haug2026efficient} and improving the techniques used to compute them~\cite{haug2023quantifying,lami2023nonstabilizerness,tarabunga2024nonstabilizerness,sinibaldi2025non}. On the other hand, there is a growing interest in different fields of application, including many-body quantum physics~\cite{liu2022many, tarabunga2023many} and variational quantum algorithms~\cite{spriggs2025quantum, capecci2025role}, where the interconnection with other quantum resources~\cite{gu2025magic, fux2024entanglement} and properties~\cite{leone2023nonstabilizerness} is particularly relevant.

In this work, we employ a GNN model to predict the stabilizer $2$-R\'enyi Entropy~\cite{lipardi2026nonstabilizerness}. Section~\ref{sre} introduces this measure of magic, while Section~\ref{gnn} describes the GNN approach.

\subsection{Stabilizer Rényi Entropy}\label{sre}
For a pure $n$-qubit quantum state $\ket{\psi}$, the \textit{stabilizer $2$-R\'enyi entropy}~\cite{leone2022stabilizer} is defined as:
\begin{equation}
    M_{2} (\ket{\psi})= - \ln \sum_{P\in \mathcal{P}_n}\Xi^{2}_P(\ket{\psi}) -\ln(2^n) \label{sre_formula}
\end{equation}
\noindent where $\Xi_P(\ket{\psi})= \frac{1}{2^n}\bra{\psi}P\ket{\psi}^2$ and $\mathcal{P}_n$ denotes the set of $n$-qubit Pauli strings. We note that $M_2$ is a nonstabilizerness monotone ~\cite{leone2024stabilizer}. 

In general, the computational cost of evaluating $M_2$ using Equation~\ref{sre_formula} scales exponentially with the number of qubits $n$~\cite{leone2022stabilizer}, as it requires estimating $4^n$ expectation values corresponding to all possible Pauli strings.
Methods based on Matrix Product States~\cite{haug2023quantifying,lami2023nonstabilizerness,lami2024unveiling,tarabunga2024nonstabilizerness}, and more in general Tensor Networks, offer a framework to compute $M_2$ efficiently in the number of qubits but in running time that scales as a polynomial of degree four in the bond dimension~\cite{haug2023quantifying}. 
These methods represent a powerful theoretical framework, but with practical limitations to low bond dimensions~\cite{tarabunga2023many, lami2023nonstabilizerness}. A recent alternative approach leverages supervised learning to generalize from computationally tractable states to larger or more complex instances~\cite{mello2025retrieving, lipardi2025study,lipardi2026nonstabilizerness}. In particular, GNNs have been proposed to estimate $M_2$, trading exact computation for fast and scalable predictions. It is important to note that learning‑based techniques are not meant to compete with tensor‑network approaches, but are complementary. For example, GNN models could be used to exploit tensor‑network computations on tractable instances and use them to extrapolate to more challenging regimes.

\subsection{Stabilizer Rényi Entropy Estimation using Graph Neural Networks} \label{gnn}
Estimating the magic of PQCs during the design PQCs is in principle unfeasible, due to the computational cost related to computing $M_2$. However, the Graph Neural Network (GNN) proposed in~\cite{lipardi2026nonstabilizerness}, enables fast $M_2$ predictions after a one-time training procedure. 

This GNN model manages to capture the relevant features of PQCs using a graph-based circuit representation, which encodes them as directed acyclic  graphs. Here, nodes correspond to quantum gates and edges represent the flow of quantum information across qubits. This structural representation enables the model to leverage message-passing layers to capture local and global correlations that are intrinsic to the magic of quantum circuits. Unlike traditional machine learning models that rely on fixed-length feature vectors~\cite{lipardi2025study}, the GNN demonstrates improved generalization capabilities for out-of-distribution circuits with higher qubit counts and gate depths than those encountered during training~\cite{lipardi2026nonstabilizerness}. By circumventing the exponential scaling of exact $M_2$ computations, the GNN model provides a computationally efficient pathway for the real-time PQC design. Section~\ref{gnn_performance} provides the details of the GNN training and assess the its performance.

\section{Quantum Architecture Search}\label{qas}
Quantum Architecture Search (QAS) aims to design a PQC that optimizes the given objective function providing hybrid quantum-classical algorithms customized on the available quantum hardware. The problem is encoded into an objective function defined on $n$-qubit PQCs with a fixed number of qubits $n$. QAS involves designing the PQC topology, which consists of an ordered sequence of quantum gates, their respective positions, and the corresponding angle parameters.
Several techniques have been proposed for QAS~\cite{zhang2022differentiable,zhang2023evolutionary, chivilikhin2020mog,he2022quantum, he2025self, kuo2021quantum, ostaszewski2021reinforcement, patel2024curriculum, wang, lipardi2025quantum}, including differentiable quantum architecture~\cite{zhang2022differentiable}, evolutionary algorithms~\cite{zhang2023evolutionary, chivilikhin2020mog}, reinforcement learning~\cite{kuo2021quantum, ostaszewski2021reinforcement, patel2024curriculum}, and Monte Carlo Tree Search (MCTS)~\cite{wang, lipardi2025quantum}. 

In this work, we focus on the Progressive Widening enhanced Monte Carlo Tree Search (PWMCTS) technique~\cite{lipardi2025quantum}, whose gradient-free approach and robustness to hyperparameter variation make it particularly suitable for the magic-informed approach. PWMCTS formulates the QAS problem as a single-player game over a search tree, allowing it to exploit the extensive previous research available on MCTS for game playing, including breakthrough approaches such as AlphaGo~\cite{silver2016mastering}.
Section~\ref{mcts} introduces the general MCTS, while Section~\ref{pwmcts} describes the PWMCTS technique~\cite{lipardi2025quantum}, which is specifically designed for the QAS problem.

\subsection{Monte Carlo Tree Search} \label{mcts}
MCTS is a best-first search method~\cite{coulom2006efficient} whose basic implementation does not require any domain-specific knowledge. 
MCTS is based on a randomized exploration of the search space. Using the results of previous explorations, MCTS gradually builds up a search tree in memory, and successively becomes better at accurately estimating the values of the most promising actions. It consists of four strategic steps~\cite{chaslot}, repeated as long as there is time left. (1) In the \textit{selection} step the tree is traversed from the root node downward until a state is chosen, which has not been stored in the tree. (2) Next, in the \textit{roll-out} step, actions are randomly chosen until a terminal state is reached. (3) Subsequently, in the \textit{expansion} step one or more states encountered along the roll-out are added to the tree. (4) Finally, in the \textit{backpropagation} step, the reward is propagated back along the previously traversed path up to the root node, where node statistics are updated accordingly. 
MCTS grows its search tree gradually by executing the four steps described above. Such an iteration is called a full simulation. 

The selection strategy used in this work is the UCB~\cite{auer2002finite}, referred to as UCT in the context of Monte Carlo Tree Search, which balances exploitation and exploration of the search~\cite{kocsis}. 
Given a state $s$ and the set of all possible actions $\mathcal{A}$, the MCTS takes the action $a^*$ with the highest $UCT$ value
\begin{equation*}
a^* = \arg \max_{a \in \mathcal{A}} \, UCT(s, a)   
\end{equation*}
\begin{equation}
UCT(s, a) = \frac{Q_{(s,a)}}{N_{(s,a)}}+ c\, \sqrt{\frac{\ln N_s}{N_{(s,a)}}} \label{uct}
\end{equation}
where $N_{(s,a)}$ is the number of times the agent took the action $a$ from the state $s$, $N_s = \sum_{a\in \mathcal{A}} N_{(s,a)}$ is the total number of times the agent visited the state $s$, and $Q_{(s,a)}$ is the cumulative reward the agent gained by taking the action $a$ from the state $s$. The constant $c$ is a parameter that controls the degree of exploration, captured by the second term of Eq.~\ref{uct} versus the exploitation captured by the first term. Here, $c$ has been set to $0.4$ based on~\cite{lipardi2025quantum}.

\subsection{Quantum Circuit Design using Progressive Widening Monte Carlo Tree Search}\label{pwmcts}
In PWMCTS~\cite{lipardi2025quantum}, each node corresponds to a $n$-qubit quantum circuit and each move corresponds to a specific modification of it. The search starts from a PQC with a Hadamard gate applied on each qubit, previously initialized to $\ket{0}$, which allows to start from a non-classical state. PWMCTS explores the search space by sampling from four classes of allowed actions:

\begin{enumerate}
    \item \textit{add} (A) a random gate on a random qubit at the end of the circuit;
    \item \textit{swap} (S) a random gate in the circuit with a new one;
    \item \textit{delete} (D) from the circuit a gate at random position;
    \item \textit{change} (C) the angle parameter $\theta_i$ of a randomly chosen parameterized gate in $\theta_i+\epsilon$, where $\epsilon \sim \mathcal{N}(0, \Delta \theta)$ is sampled from a normal distribution with mean zero and standard deviation $\Delta \theta$.
\end{enumerate}

MCTS chooses between these four classes by sampling from a probability mass distribution $p=(p_A, p_S, p_C, p_D)$,
where $p_A$, $p_S$, $p_C$, $p_D$ correspond to the probabilities of choosing the respective action. Quantum gates are sampled from the universal gate set $G=( CX, R_x, R_y, R_z)$. This technique is inspired by a framework~\cite{franken2022quantum} proposed for an evolutionary strategy on quantum circuits.  
PWMCTS is equipped with a progressive widening technique~\cite{chaslot} to restrict the infinite discrete number of moves allowed from a state $s$ to a finite number of moves:
\begin{equation}
    k_s=\lceil \beta_{PW} N_s^{\alpha_{PW}} \rceil    \label{progressive_widening}
\end{equation}
where $\alpha_{PW} \in \; ]0,1]$ and $\beta_{PW} >0$ are hyperparameters of PWMCTS. All hyperparameters, including $p$, $\Delta \theta$, $\alpha_{PW}$, and $\beta_{PW}$ have been fixed according to the previous work \cite{lipardi2025quantum}.
Because we are interested in the whole sequence of gates, an action-by-action search is employed to distribute the search time along all the levels of the tree~\cite{baier2012nested,schadd2012single}. Once a level of the tree is sufficiently explored, the MCTS commits to the best action.

\section{Magic-Informed Progressive Widening enhanced Monte Carlo Tree Search}\label{magic_informed_pwmcts}
This section describes the proposed magic-informed PWMCTS technique, in which GNN-based estimates are used to induce a magic-based bias in the tree search. 
Section~\ref{subsec:magic_informed_PWMCTS} details the magic-informed variant of the baseline PWMCTS, which operates at two stages, namely expansion and selection. 
Section~\ref{gnn_performance} analyzes the quality of the GNN estimates for both the $M_2$ prediction and ranking tasks, which are essential for the effectiveness of the magic-informed expansion and selection, respectively. Finally, Section~\ref{subsec:testbed} provides a procedure to design quantum circuits with a fixed number of gates and different levels of magic.

\subsection{Magic Expansion and Selection Strategies}\label{subsec:magic_informed_PWMCTS}
The magic-informed PWMCTS can be divided into two classes: the high-magic class, which biases the search toward PQCs with high $M_2$ to obtain solutions that are harder to simulate classically, and the low-magic class, which biases the search toward PQCs with low $M_2$ to obtain solutions that are easier to simulate. The magic-informed PWMCTS technique extends the baseline PWMCTS by introducing two independent magic-based biases: the magic progressive widening in the expansion step and the magic UCT as a selection strategy.  
We emphasize that $M_2$ estimates are exclusively used to bias the growth of the search tree, and the evaluation of the quality of candidate PQCs relies on a problem-dependent reward.

The first bias modifies the progressive widening mechanism that defines the expansion strategy, see Equation~\ref{progressive_widening}. In the magic progressive widening (magic PW), the number of child nodes $k_s$ for a state $s$, as defined by Equation~\ref{progressive_widening}, is rescaled by a multiplicative factor $\gamma$.
At each expansion step, a batch of $\bar{k}_s = \gamma k_s$ candidate circuits is generated, with $\gamma>1$. These circuits are evaluated in parallel using a batched GNN forward pass to obtain their predicted $\hat{M}_2$ values. The top-$k_s$ candidates are then retained in the search tree, while the remaining $\bar{k}_s - k_s$ candidates are discarded. The hyperparameter $\gamma$ is fixed to $3$ in order to keep limited the computational overhead of the GNN prediction time. We note that progressive strategies have been proven to be particularly effective in the context of MCTS for game playing~\cite{chaslot}, where domain-specific heuristics are used to guide the exploration process and improve performance.

The second bias is a PUCT-like selection strategy~\cite{rosin2011multi} that introduces a magic-based multiplicative factor to the exploration term of the standard UCT, reported in Equation~\ref{uct}. The resulting magic UCT formula is:
\begin{equation}
    UCT_{M_2}(s, a) = \frac{Q_{(s,a)}}{N_{(s,a)}}+ c \cdot \, p_{M_2}\, \sqrt{\frac{\ln N_s}{N_{(s,a)}}} 
    \label{eq:magic_ucb}
\end{equation}
where $p_{M_2} \equiv p_{M_2}(s,a)$ is the normalized $M_2$ prediction of the GNN model on the circuit obtained from state $s$ after applying action $a$. For the bias toward high-magic state we use $p_{M_2}=\frac{\hat{M_2}}{M_2^{\max}}$, while for the bias to low-magic states $1-\frac{\hat{M_2}}{M_2^{\max}}$. The maximum value $M_2^{\max} = \ln\left(\frac{2^n + 1}{2}\right)$, derived in~\cite{leone2022stabilizer} as a function of the number of qubits $n$, ensures the normalization of $p_{M_2}$. This value is computed once and cached at each node. There are two main differences between the proposed magic UCT approach and the plain PUCT selection strategy. First, the exploration term scales differently with the visit counts, as in PUCT deviates from the standard UCT formulation and scales with $\frac{\sqrt{N_s}}{N_{(s,a)}}$~\cite{rosin2011multi, silver2016mastering}. Second, the multiplicative factor in PUCT is usually a probability distribution over all possible actions $a$. In contrast, although $p_{M_2} \in [0,1]$ for all nodes, it does not form a normalized distribution for a given state $s$: $\sum_a p_{M_2}(s,a) \neq 1$. We note that PUCT-like selection strategies have been successfully used in AlphaGo~\cite{silver2016mastering}, where they incorporate prior policy information captured by a deep-learning model in the tree search.

Magic PW and magic UCT modify the baseline PWMCTS at two different stages and are therefore independent from each other. The resulting magic-informed PWMCTS includes both contributions. The magic PW ensures that the search tree is populated with PQCs exhibiting higher (or lower) $M_2$, while the magic UCT selection strategy biases the traversal towards subtrees characterized by similar magic properties. As a result, the hybrid approach simultaneously guides both the expansion and selection strategies using the estimations $\hat{M}_2$.

\subsection{Graph Neural Network Model}\label{gnn_performance}
The Graph Neural Network employed in this work adopts the same architecture, hyperparameters, and training dataset as those described in~\cite{lipardi2025quantum}. Specifically, the dataset used here is referred to as Random Quantum Circuit dataset, with circuits from $2$ to $6$ qubits and total gates ranging from $1$ to $99$. Any reference to the GNN model in the remainder of the paper refers to this implementation.

\begin{figure}[h]
    \centering
    \includegraphics[width=1\linewidth]{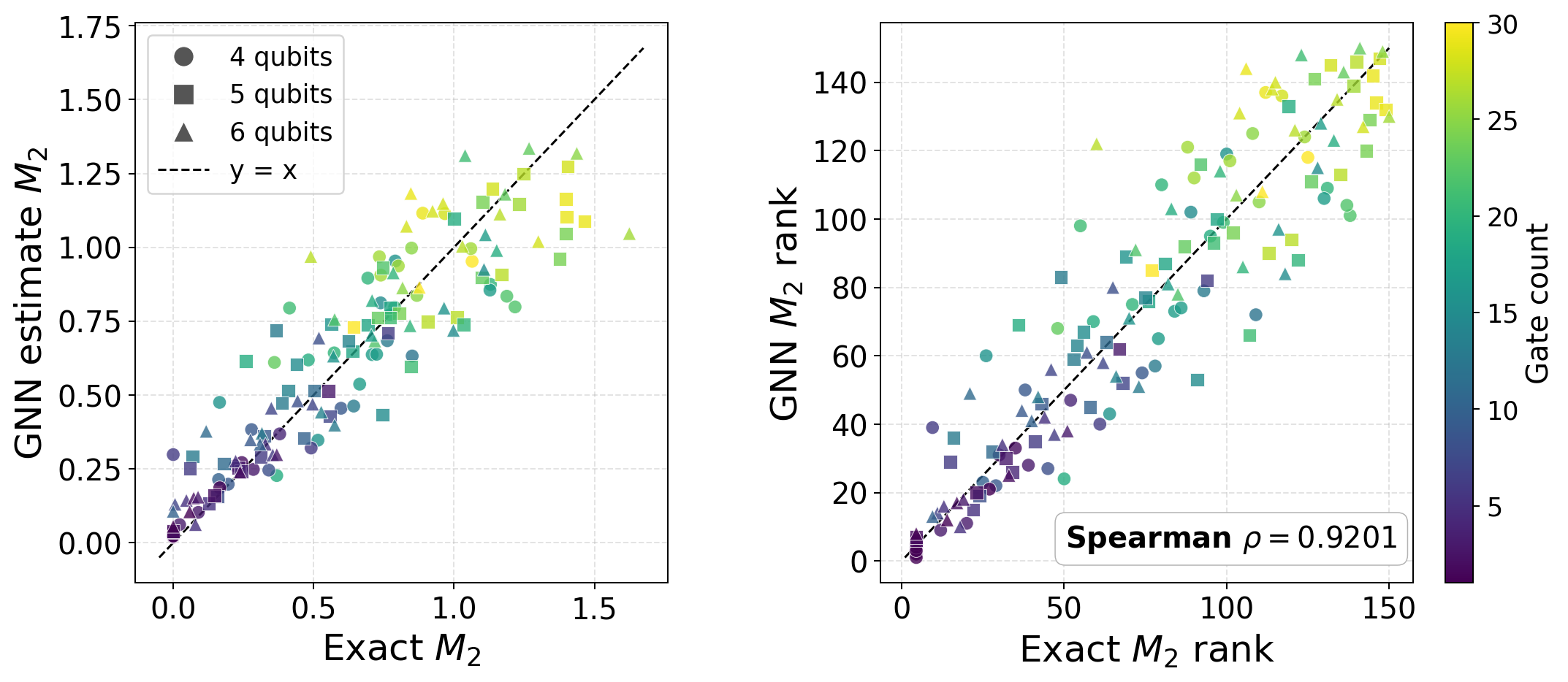}
    \caption{Graph Neural Network performance. On the left, the performances in terms of $M_2$ estimations, on the right, in terms of Spearman's ranking correlation. }
    \label{fig:gnn_spearman}
\end{figure}

Here, we evaluate the $\hat{M_2}$ estimations provided by the GNN model. In particular, we focus on the PQC sixe that is typically explored by PWMCTS, based on its constraints and hyperparameters. For example, the maximum circuit depth, probability distribution over the four possible actions, and the progressive widening technique, typically restrict the search space to PQCs with total gates between $1$ and $30$. Because the magic-informed PWMCTS variants use the GNN at two stages, as a ranking model for the magic progressive widening and as a $M_2$ estimator for the Magic UCT, the performance is evaluated in terms of $M_2$ estimations and on the widely used Spearman's rank correlation coefficient $\rho_S$~\cite{spearman1961proof}. 
Figure~\ref{fig:gnn_spearman} reports the GNN estimations on a set of $150$ of random quantum circuits, with $50$ circuits for each qubit count $n\in\{4,5,6\}$. For each circuit, the number of gates is drawn uniformly from the interval $ [ 1,30]$, and individual gates are sampled from the set $(CNOT, H, R_X, R_Y, R_Z)$. The exact $M_2$ is computed for each circuit, alongside the corresponding GNN prediction. 
 
\subsection{Generation Procedure of Target States}
\label{subsec:testbed}
\begin{figure}[ht]
    \centering
    \includegraphics[width=1 \linewidth]{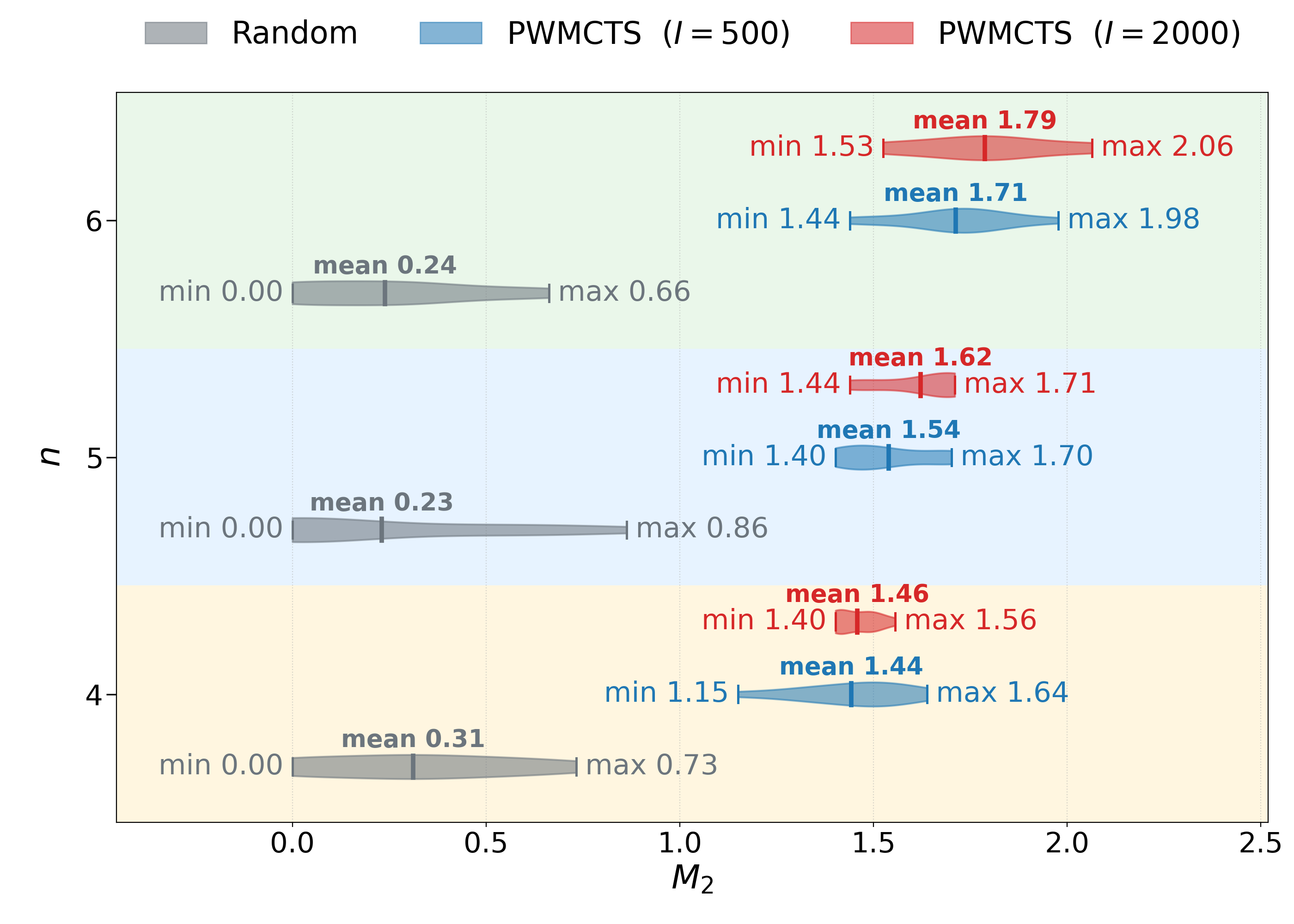}
    \caption{Generation procedure of the target states. For each value of the number of qubits $n$, the plot shows the $M_2$ value distribution of the generated random quantum circuits (grey), of the resulting circuits optimized in $M_2$ using PWMCTS with a number of iterations fixed to $500$ (blue), and $2000$ (red). }
    \label{fig:sre_distribution}
\end{figure}

We generate target states with a number of qubits spanning $n \in \{4,5,6\}$, with three representative magic levels for each size. For each value of $n$, we proceed as follows. First, we generate a pool of $10$ random circuits, by uniformly sampling $20$ gates from the gate set $\{\text{CNOT}, H, S, T\}$, each applied on a random qubit. Since these circuits exhibit low magic (see the gray violin plots in Figure~\ref{fig:sre_distribution}), we select the circuit with the highest $M_2$ as the \emph{low-magic} representative, and the one with the lowest magic as the \emph{working circuit} that is used as a starting circuit to design two more target circuits with higher $M_2$. 
Next, we apply PWMCTS to increase the magic level while keeping the total number of gates fixed. In this setup, PWMCTS uses $M_2 / M_2^{\max}$ as reward function, and the action probabilities are set to $p=(0,40,60,0)$, effectively disabling the \emph{add} and \emph{delete} actions in favor of \emph{swap} and \emph{change}. We note that this reward function takes values in $[0,1]$.
For each $n$, we perform 10 independent PWMCTS runs, where we initialize the root node in the selected \emph{working circuit}. We consider two iteration budgets, $I=500$ and $I=2000$, producing two sets of optimized circuits with increasing magic. The circuit with the lowest $M_2$ among those obtained with $I=500$ is selected as the \emph{medium-magic} target circuit, while the circuit with the highest $M_2$ among those obtained with $I=2000$ is selected as the \emph{high-magic} target circuit. 

This procedure generates nine target quantum circuit differing in terms of number of qubits $n$ and magic $M_2$.
Figure~\ref{fig:sre_distribution} illustrates the distribution of $M_2$ value for the initial random quantum circuits generated for each value of $n$, together with the resulting circuit obtained by applying PWMCTS on the random quantum circuit with the lowest $M_2$. As shown in the figure, this initial circuit consistently corresponds to a stabilizer state. The reason for starting from the circuit with the lowest $M_2$ is twofold. First, it is different from the circuit with the highest $M_2$ value, which is already part of the target circuit set. Second, it showcases the capabilities of PWMCTS in the circuit optimization task, where even with low number of iterations and starting from a stabilizer state it successfully designs a circuit with higher magic using the limited resources given by the number of total gates. 

\begin{table}[h]
\centering
\caption{Target Circuits Summary}
\label{tab:target_circuits}
\begin{tabular}{c c c c c c}
\hline
$n$ & Magic Level & CNOT &T Count& $M_2$ \\
\hline
4 & Low    & 3 &  6 &0.73 \\
4 & Medium & 3 &  5 &1.15\\
4 & High   & 6 &  6 &1.56\\
5 & Low    & 3 &  4 &0.86 \\
5 & Medium & 4 &  5 &1.40 \\
5 & High   & 5 &  6 &1.71 \\
6 & Low    & 4 &  4 &0.66 \\
6 & Medium & 5 &  8 &1.44 \\
6 & High   & 3 &  9 &2.06 \\
\hline
\end{tabular}
\end{table}

It is important to note that the PWMCTS employed for generating target circuits is considerably different from the PWMCTS used for the QAS context. They are initialized on different circuits, share different hyperparameters, and most importantly they work on different universal gate sets. This distinction is essential to avoid structural biases that could artificially simplify the problem. 
Table~\ref{tab:target_circuits} summarizes the nine target circuits designed for the quantum state approximation problem. We emphasize that PWMCTS can solve the nontrivial task of placing non-Clifford T gates in appropriate positions to increase magic. For example, low- and high-magic $4$-qubit circuits may share the same number of T gates, yet the circuit produced by PWMCTS achieves an $M_2$ value more than twice as large. This approach enables the successful design of target circuits spanning a wide range of magic levels.

\section{Experimental Results}\label{experiments}
This section presents a set of experiments to evaluate the performance of the proposed magic-informed PWMCTS technique against the standard PWMCTS. 
Section~\ref{subsec:quantum_chemistry} presents the results on the ground-state energy problem for three different molecules. Section~\ref{subsec:quantum_state_approximation} focuses on the more general quantum state approximation problem, where the target solutions are the nine random quantum circuits generated with different magic levels, as detailed in~\ref{subsec:testbed}. The experiments in these two sections are divided in two main parts. The first analyzes the ability of the proposed technique to control the magic levels within the tree search and in the final PQC. The second analyzes the performance in terms of solution quality, which are the energy and fidelity for the ground-state energy and quantum state approximation respectively. To isolate the contribution of the magic-based components, we conduct an ablation study analyzing the independent effects of the magic PW and magic UCT strategies. 
Finally, Section~\ref{structures} analyzes the impact of the magic-informed PWMCTS on the structure of the PQCs designed.

\subsection{Ground-state Energy Problem} \label{subsec:quantum_chemistry}
The ground-state energy problem is fundamental in the field of quantum chemistry. Any molecule is described by an Hamiltonian that encodes the electronic structure of the system. The problem consists of determining the eigenvector (ground state) corresponding to the lowest eigenvalue of this Hamiltonian (minimum energy). Here, we consider three molecules: hydrogen (H$_2$), water (H$_2$O), and lithium hydride (LiH).
Table~\ref{tab:molecules} summarizes the specifics of these molecules, including the number of active electrons $n_e$, active orbitals $n_o$, the number of qubit $n$ required for the simulation, and the molecular geometry, which specifies the position of each atom in three-dimensional space. These three molecules have been chosen according to the baseline work~\cite{lipardi2025quantum} as they span various system sizes and complexities. 

\begin{table}[ht]
\centering
\caption{Molecule Structures}
\label{tab:molecules}
\begin{tabular}{l c c c c}
\hline
Molecule & $n_e$ &$n_o$ & $n$ & Geometry \\
\hline
H$_2$   & 2 & 2 & 4  & $(0,0,\pm0.6614)$ \\

H$_2$O  & 4 & 4 & 8  & H: $(0,0,0)$, $(3.36,0,0)$; \\
        &     &   &    & O: $(1.63,0.86,0)$ \\

LiH     & 2 & 5 & 10 & $(0,0,0)$, $(0,0,2.97)$ \\
\hline
\end{tabular}
\end{table}
The following study is performed under a fixed  computational budget, measured in terms of the number of PWMCTS iterations $I$, which correspond to the total number of circuit evaluations~\cite{lipardi2025quantum}. In this application, we set $I=1000$ for H$_2$, $I=50000$ for H$_2$O, and $I=5000$ for LiH. We perform 10 independent runs for each PWMCTS variant, for both the low-magic and high-magic class as defined in Section~\ref{magic_informed_pwmcts}.

\begin{figure}[h]
    \centering
    \includegraphics[width=\linewidth]{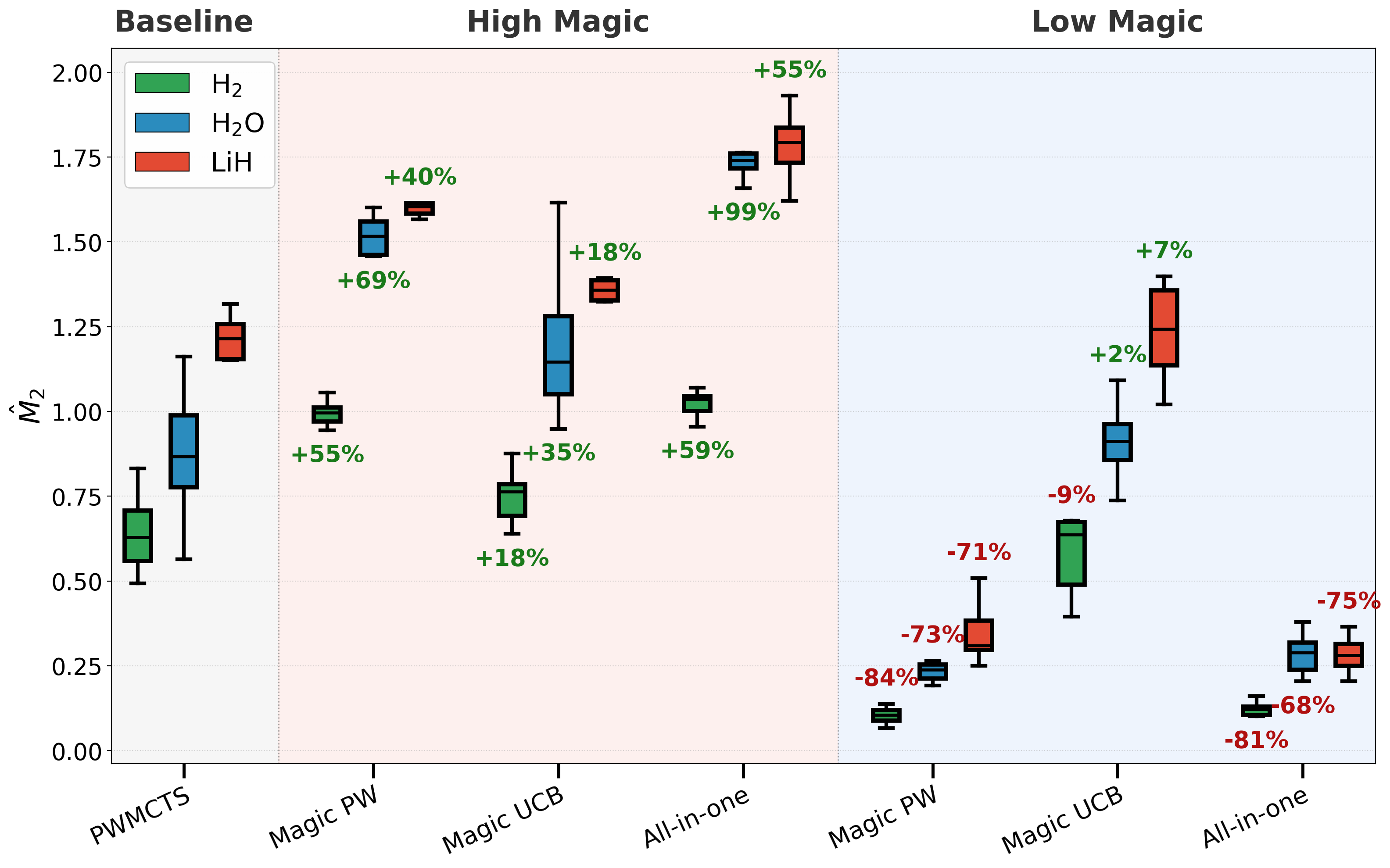}
    \caption{Average GNN estimations $\hat{M}_2$ in the search tree.}
    \label{fig:avg_sre}
\end{figure}

The boxplots in Figure~\ref{fig:avg_sre} show the average GNN estimations $\hat{M}_2$ across all nodes in the search tree for each PWMCTS variant and for the three molecules. From left to right, the figure reports results for the baseline PWMCTS, the variants of the high-magic class, and of the low-magic class. Each class includes the ablation settings (along the x-axis) that isolate the independent contributions of the two magic-based bias strategies. In the figure, the horizontal black lines in the boxplots indicate the median values, while the values in percentage correspond to the relative change in the mean $\hat{M}_2$ with respect to the baseline. Overall, the magic-informed PWMCTS significantly shifts the distribution of $\hat{M}_2$, with the magic PW having a stronger impact on $\hat{M}_2$ than the magic UCT. When combined, the two strategies consistently amplify the observed shifts, leading to the largest deviations from the baseline values.

\begin{figure*}[h]
\centering
\subfloat[H$_2$]{%
    \includegraphics[width=0.30\textwidth]{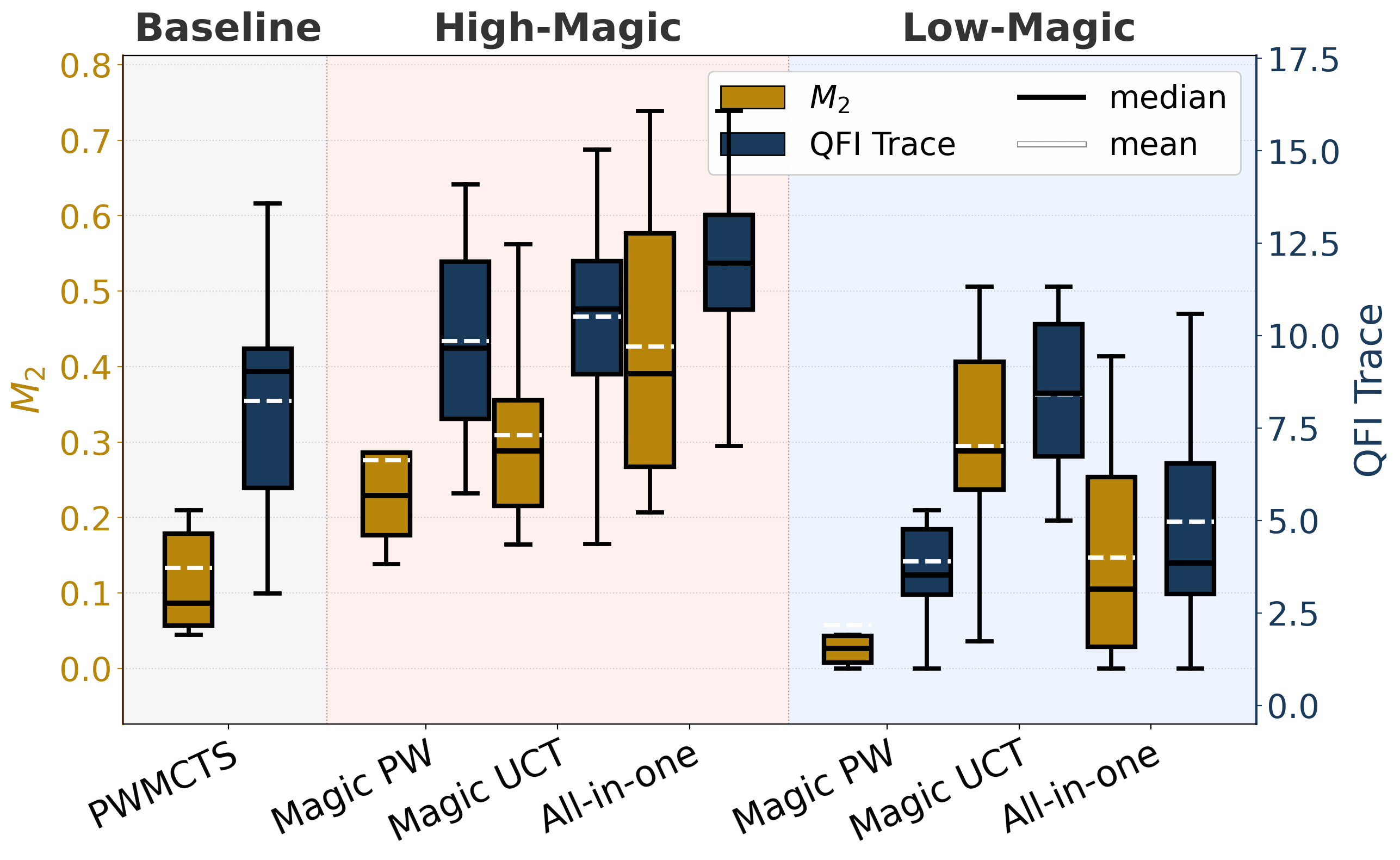}
    \label{fig:su}
}
\hspace{0.02\textwidth}
\subfloat[H$_2$O]{%
    \includegraphics[width=0.30\textwidth]{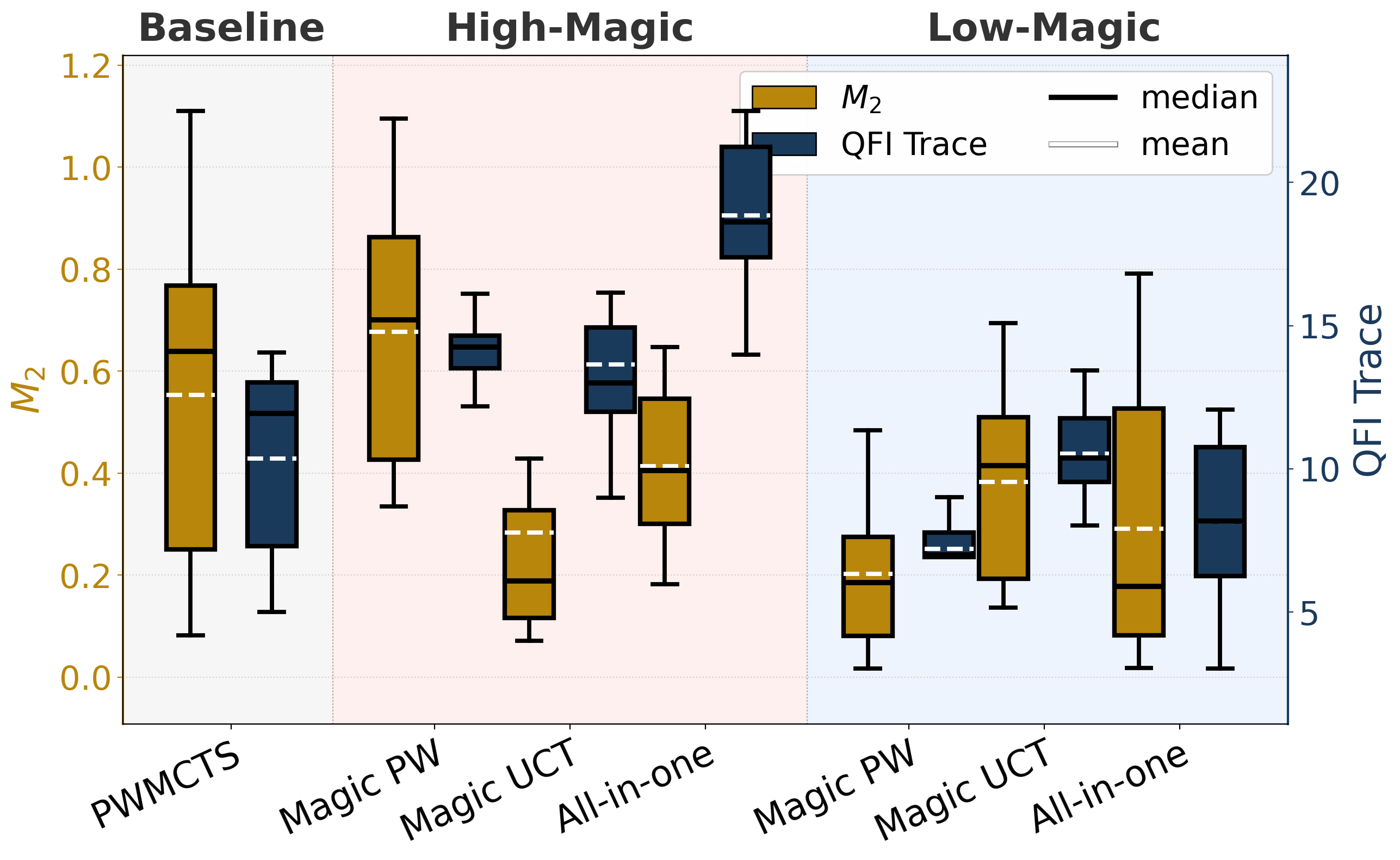}
    \label{fig:sub2}
}
\hspace{0.02\textwidth}
\subfloat[LiH]{%
    \includegraphics[width=0.30\textwidth]{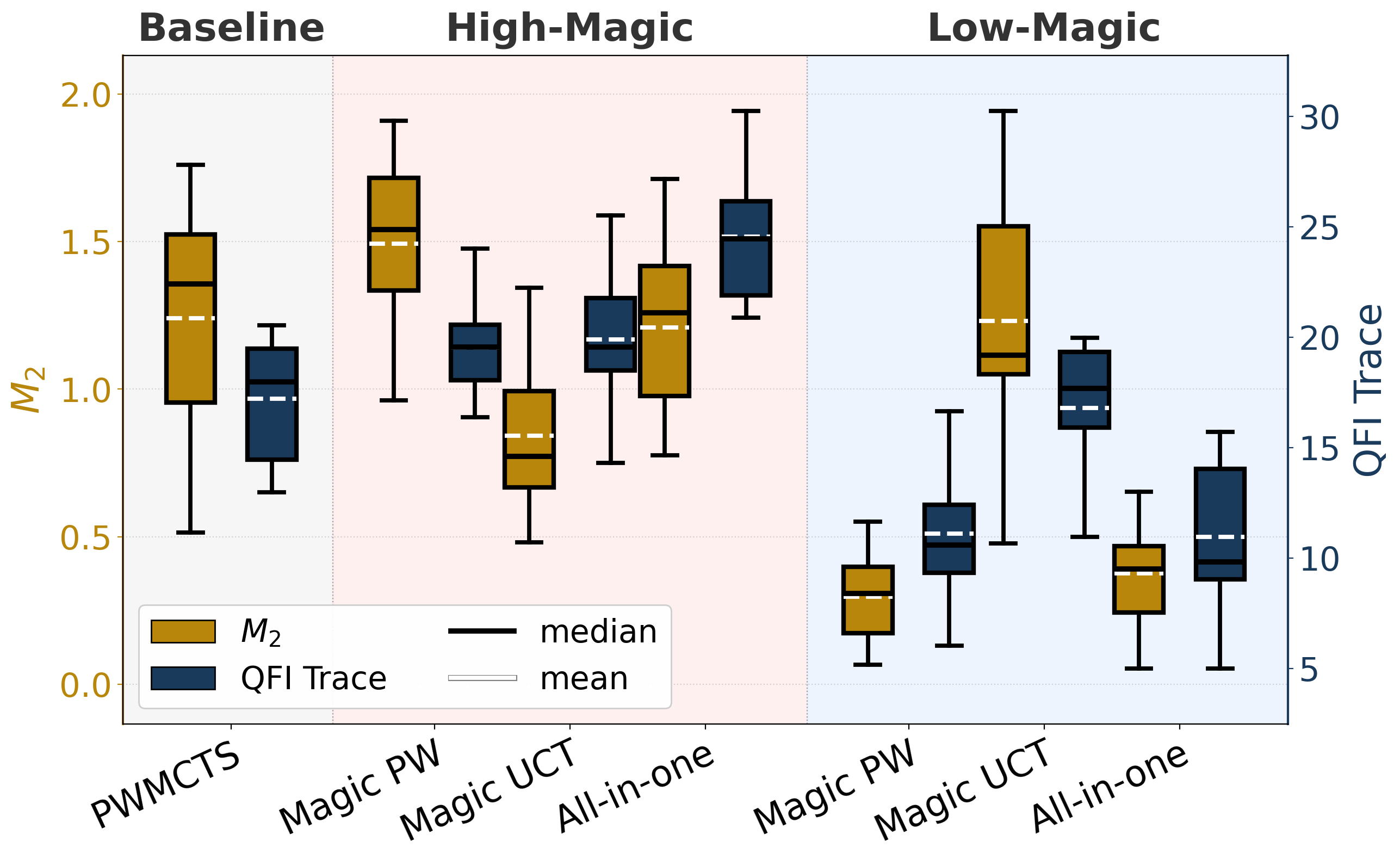}
    \label{fig:sub3}
}
\caption{Ground-state energy problem. Exact $M_2$ values and trace of the QFI matrix of the PQCs designed by each gradient-free PWMCTS variant.}
\label{fig:sre_qfi_mol}
\end{figure*}

We note that a high average $\hat{M}_2$ does not guarantee higher $M_2$ values in the final PQC. Firstly, because the reward function is generally independent on the magic. Secondly, because the effectiveness of the bias relies on the accuracy of the GNN estimations $\hat{M_2}$. As a consequence, we assess this point by computing the exact $M_2$ of the final PQC. Figure~\ref{fig:sre_qfi_mol} reports these values in the red boxplots, together with the trace of Quantum Fisher Information in blue.

\begin{figure}[h]
    \centering
    \includegraphics[width=1 \linewidth]{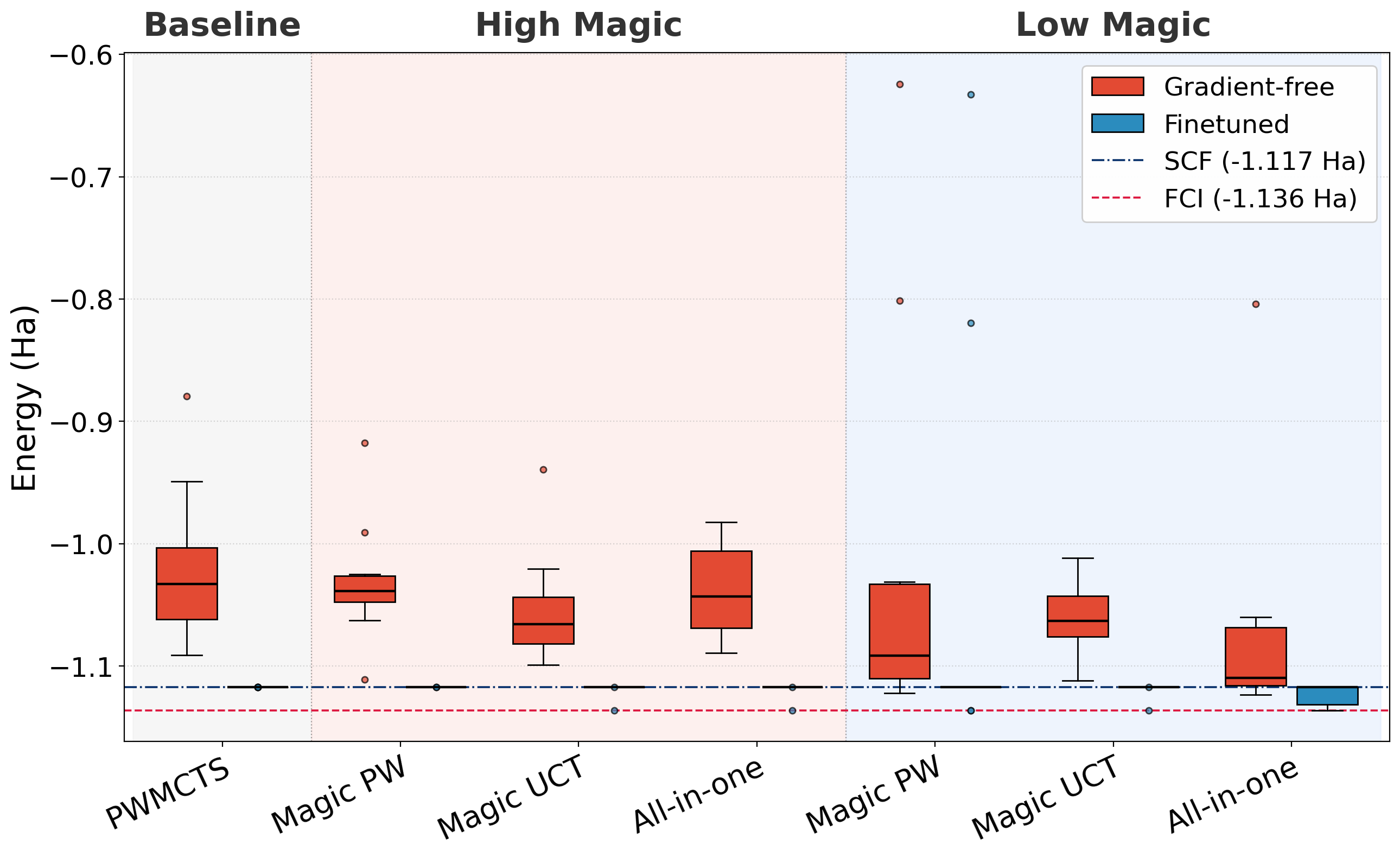}
    \caption{H$_2$. Energy achieved by each PWMCTS variant.}
    \label{fig:h2}
\end{figure}

\begin{figure}[h]
    \centering
    \includegraphics[width=1 \linewidth]{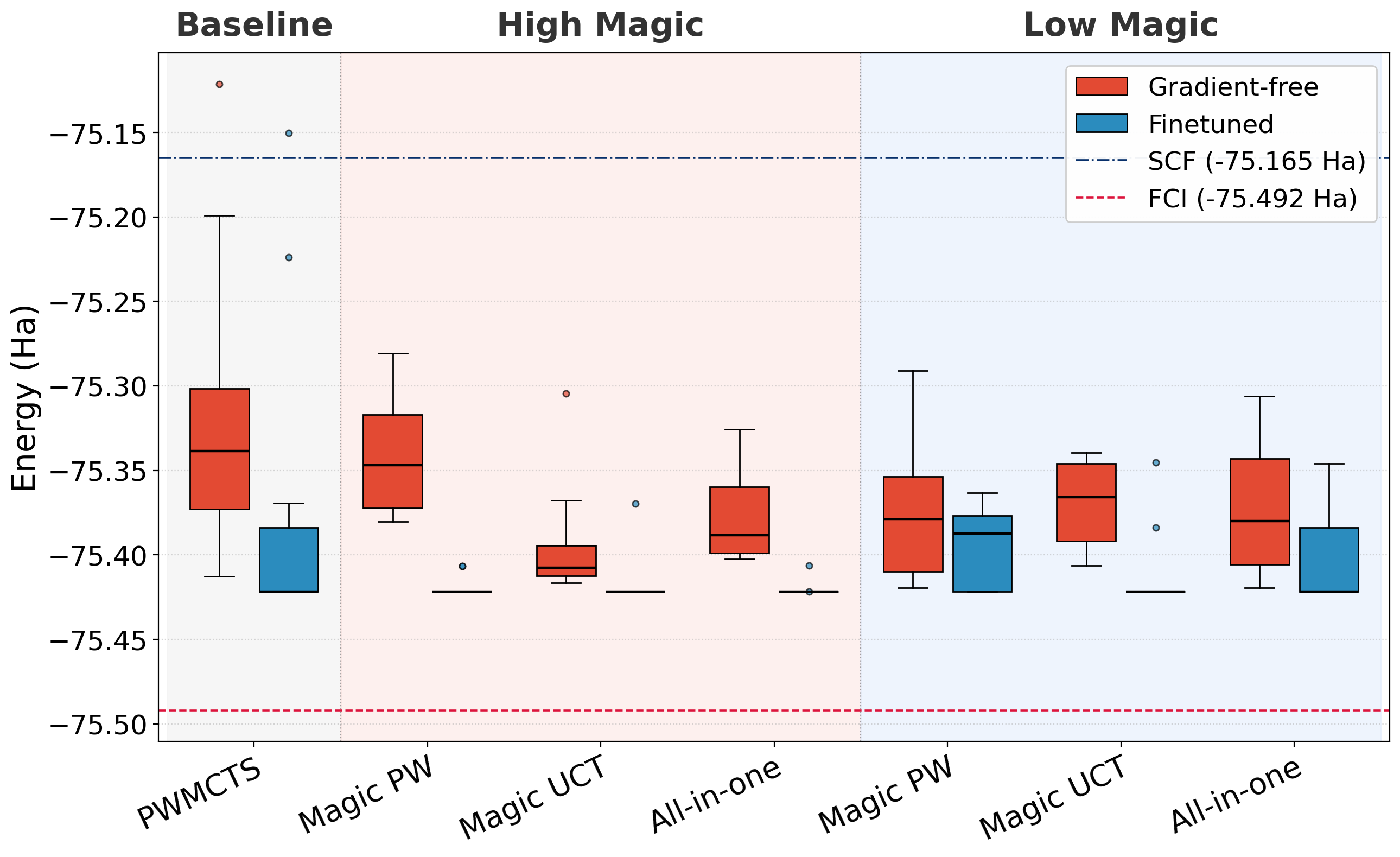}
    \caption{H$_2$O. Energy achieved by each PWMCTS variant.}
    \label{fig:h2o}
\end{figure}

\begin{figure}[h]
    \centering
    \includegraphics[width=1 \linewidth]{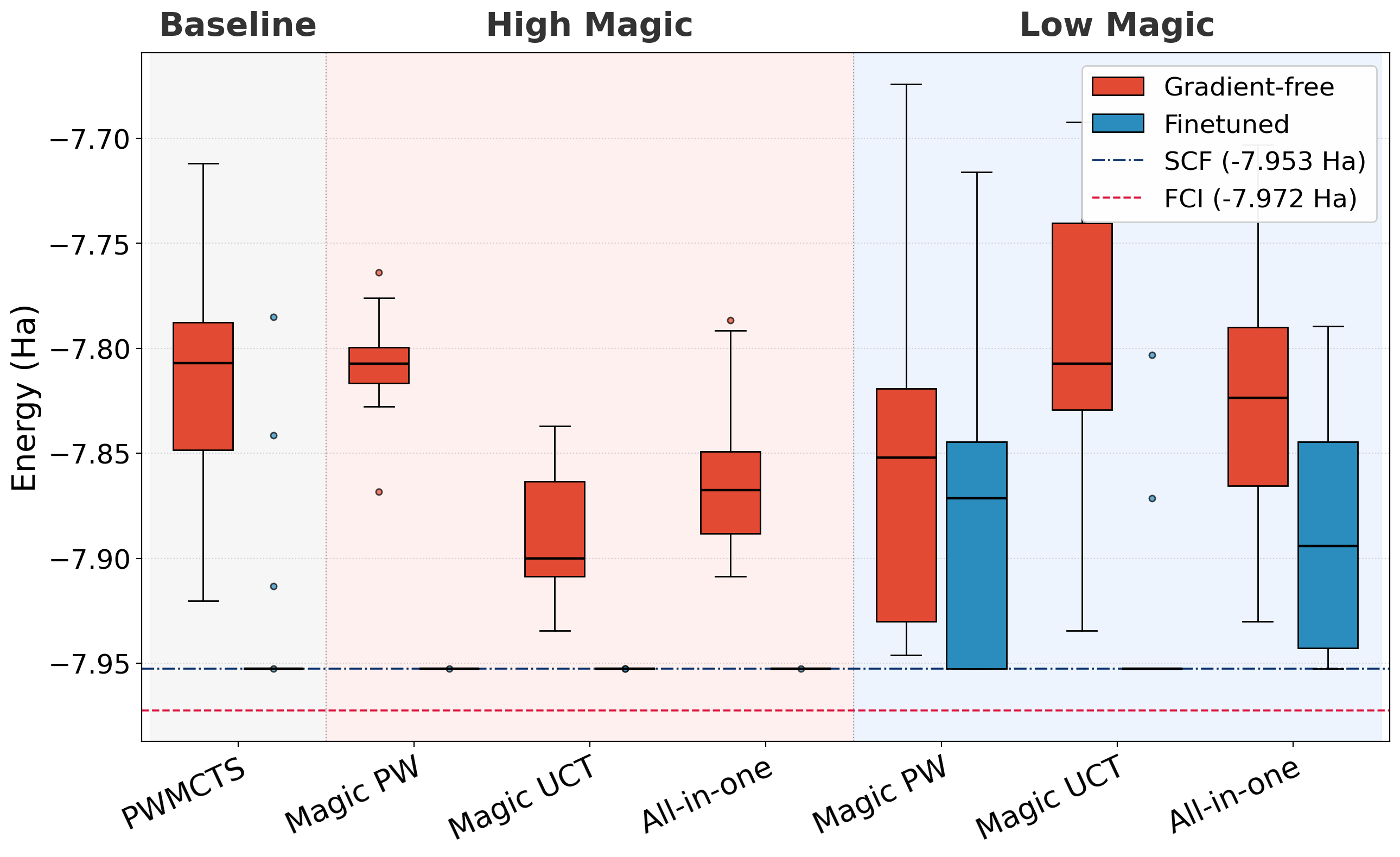}
    \caption{LiH. Energy achieved by each PWMCTS variant.}
    \label{fig:lih}
\end{figure}

The analysis of the performance based on the solution quality is illustrated in Figures ~\ref{fig:h2}, ~\ref{fig:h2o}, and~\ref{fig:lih} for the molecules of H$_2$, H$_2$O, and LiH, respectively. Orange boxplots correspond to the energies achieved by the circuits designed by the gradient-free PWMCTS variants before the parameter finetuning, whereas blue boxplots correspond to the energies obtained from the same PQCs after finetuning using the Adam optimizer~\cite{kingma2014adam}.
The horizontal lines indicate the classical benchmarks provided by the self-consistent field (SCF) and the full configuration interaction (FCI) methods. SCF is a mean-field approximation in which electrons are treated independently within an averaged potential, whereas FCI provides an exact solution by accounting for all possible electronic configurations and thus fully capturing electron correlation effects. The FCI method is exact within a fixed basis set, which according to previous works we set to STO-3G~\cite{wang, lipardi2025quantum}.

The experiments on the simplest molecule H$_2$ show that all variants manage to converge to the SCF solution. However, the magic-informed PWMCTS converge more reliably in the gradient-free phase to the solution compared to the baseline, and the combination of magic PW and magic UCT (all-in one) results in a systematic lower energy approaching to the FCI solution. This suggests that inducing the correct magic-based bias can improve the convergence of PWMCTS to the solution.  

The results on the more complex molecules of H$_2$O and LiH, show that high-magic variants achieve improved performance compared to the baseline. In this case, the increased molecular complexity requires more expressive circuit structures. On the one hand, high-magic variants provide better solutions in the gradient-free stage, and after the parameter finetuning they converge with higher reliability to the energy value $-75.445$. On the other hand, the variants belonging to the low-magic class perform similarly to the baseline, with the exception of the magic UCT, which also induces a weaker magic bias.

\subsection{Quantum State Approximation} \label{subsec:quantum_state_approximation}
In the quantum state approximation, also referred to as quantum oracle approximation or unitary compilation, the goal is finding a PQC that prepares a quantum state close to the target state. We measure the distance between two circuits in terms of fidelity, which we compute through noiseless state vector simulations. 

\begin{figure}[h]
     \centering
     \includegraphics[width=0.9\linewidth]{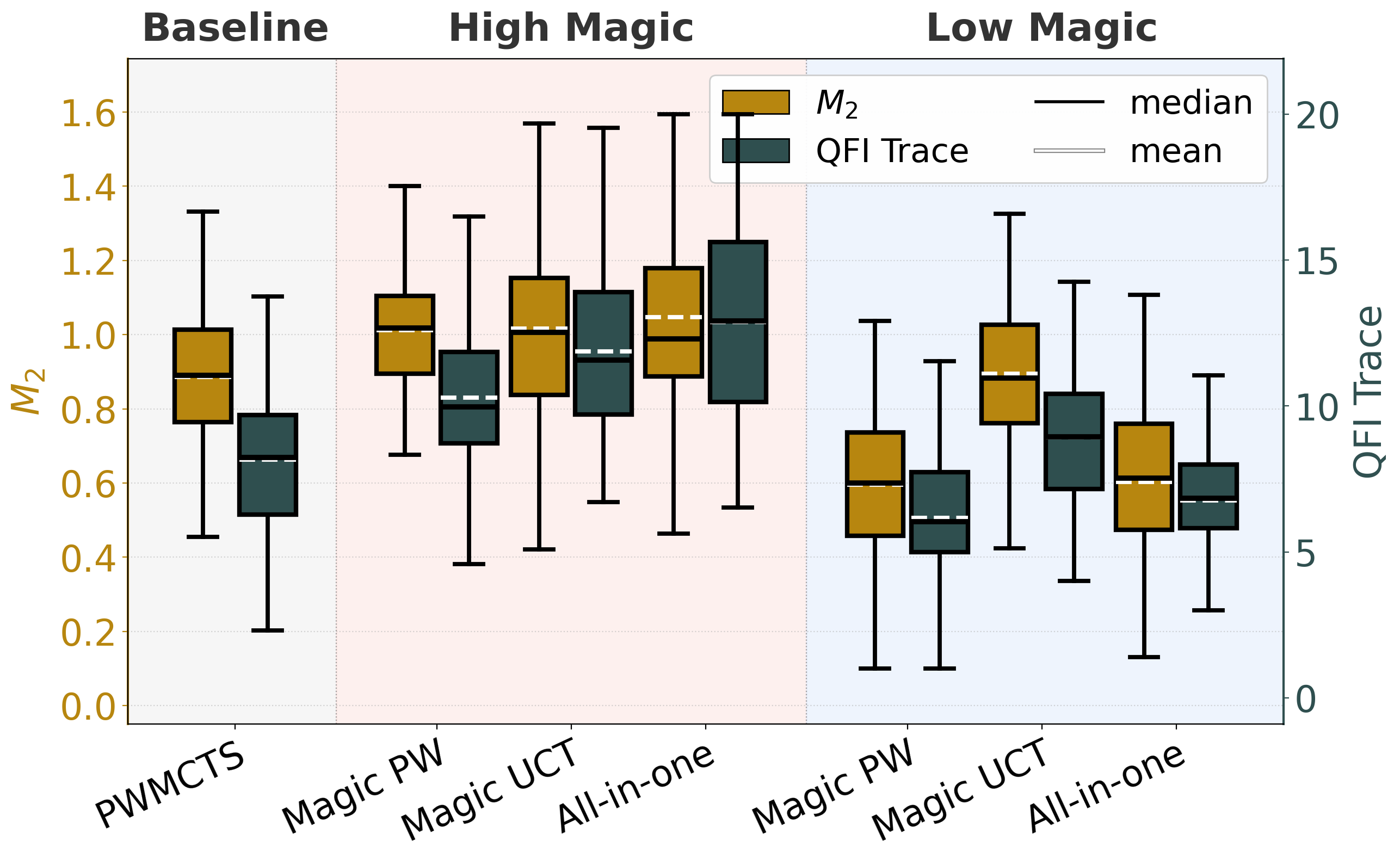}
     \caption{Exact $M_2$ values and trace of the QFI matrix of the PQCs designed by each gradient-free PWMCTS variant.}
     \label{fig:sre_qfi}
 \end{figure}

We evaluate the performance of the magic-informed PWMCTS on the set of nine target quantum states, generated following the procedure described in Section~\ref{subsec:testbed}, and summarized in Table~\ref{tab:target_circuits}. This procedure constructs a set of circuits spanning a wide range of $M_2$ values while keeping the number of gates fixed. For a fair comparison, we perform 10 independent runs for each PWMCTS variant under a fixed iteration budget $I=20{,}000$. Because the average $M_2$ estimations in the search tree exhibits similar trends to those observed in Figure~\ref{fig:avg_sre}, we focus directly on the $M_2$ on the final PQC. To avoid bias due to approximation errors in the GNN, all $M_2$ are computed exactly. Figure~\ref{fig:sre_qfi} reports the $M_2$ values of the PQCs designed by all the variants aggregated on all  nine problems. According to the results presented in the previous section, the $M_2$ values of the designed PQC are successfully biased in the desired direction by both the magic PW and the magic UCT for both high-magic and low-magic setting. 

\begin{table*}[h]
\centering
\caption{Summary of the structure of PQCs designed by magic-informed variants against the PWMCTS baseline.}
\label{tab:combined_results}
\setlength{\tabcolsep}{6pt}
\begin{tabular}{l l l l c c c}
\toprule
\textbf{} & \textbf{Problem} & \textbf{Variant} & \textbf{Class}
  & \textbf{Total gates}
  & \textbf{Parameterized gates}
  & \textbf{Adam steps} \\
\midrule
\multirow{6}{*}{\shortstack{\textbf{Ground-state}\\\textbf{Energy Estimation}}}
& \multirow{2}{*}{H$_2$ }
  & PWMCTS      & ---         & $16\pm4$  & $12\pm4$  & $122\pm33$  \\
&
  & All-in-one  & Low-Magic   & $16\pm6$  & $9\pm4$   & $118\pm56$  \\
\cmidrule(lr){2-7}
& \multirow{2}{*}{H$_2$O }
  & PWMCTS      & ---         & $23\pm3$  & $14\pm4$  & $103\pm21$  \\
&
  & All-in-one  & High-Magic  & $36\pm3$  & $30\pm3$  & $156\pm74$  \\
\cmidrule(lr){2-7}
& \multirow{2}{*}{LiH }
  & PWMCTS      & ---         & $34\pm4$  & $23\pm3$  & $163\pm34$  \\
&
  & All-in-one  & High-Magic  & $45\pm5$  & $36\pm3$  & $197\pm112$ \\
\midrule
\multirow{6}{*}{\shortstack{\textbf{Quantum State}\\\textbf{Approximation}}}
& \multirow{2}{*}{Low-Magic Circuits}
  & PWMCTS      & ---         & $19\pm4$  & $12\pm3$  & $130\pm30$  \\
&
  & Magic PW    & Low-Magic   & $16\pm3$  & $10\pm2$  & $108\pm22$  \\
\cmidrule(lr){2-7}
& \multirow{2}{*}{Medium-Magic Circuits}
  & PWMCTS      & ---         & $20\pm4$  & $14\pm3$  & $111\pm12$  \\
&
  & All-in-one  & High-Magic  & $28\pm6$  & $22\pm5$  & $109\pm42$  \\
\cmidrule(lr){2-7}
& \multirow{2}{*}{High-Magic Circuits}
  & PWMCTS      & ---         & $19\pm3$  & $12\pm3$  & $142\pm56$  \\
&
  & Magic PW    & High-Magic  & $22\pm3$  & $16\pm2$  & $122\pm23$  \\
\bottomrule
\end{tabular}
\end{table*}

\begin{figure*}[h]
    \centering
    
    \begin{subfigure}{0.95\linewidth}
        \centering
        \includegraphics[width=\linewidth]{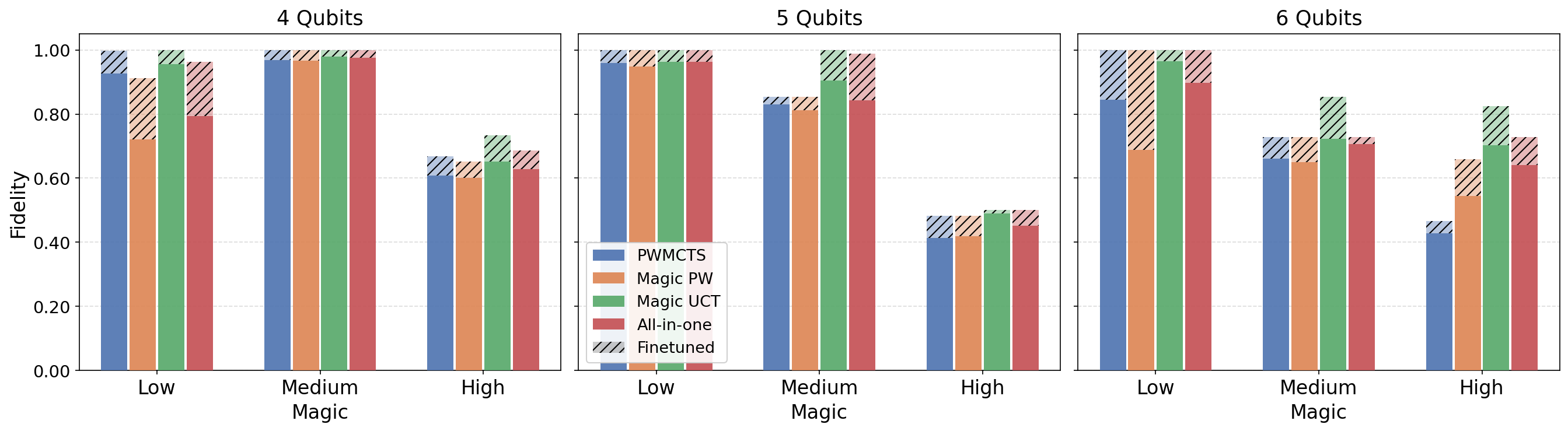}
        \caption{High-magic PWMCTS variants}
        \label{fig:fidelity_high}
    \end{subfigure}
    
    \vspace{0.5em}
    
    \begin{subfigure}{0.95\linewidth}
        \centering
        \includegraphics[width=\linewidth]{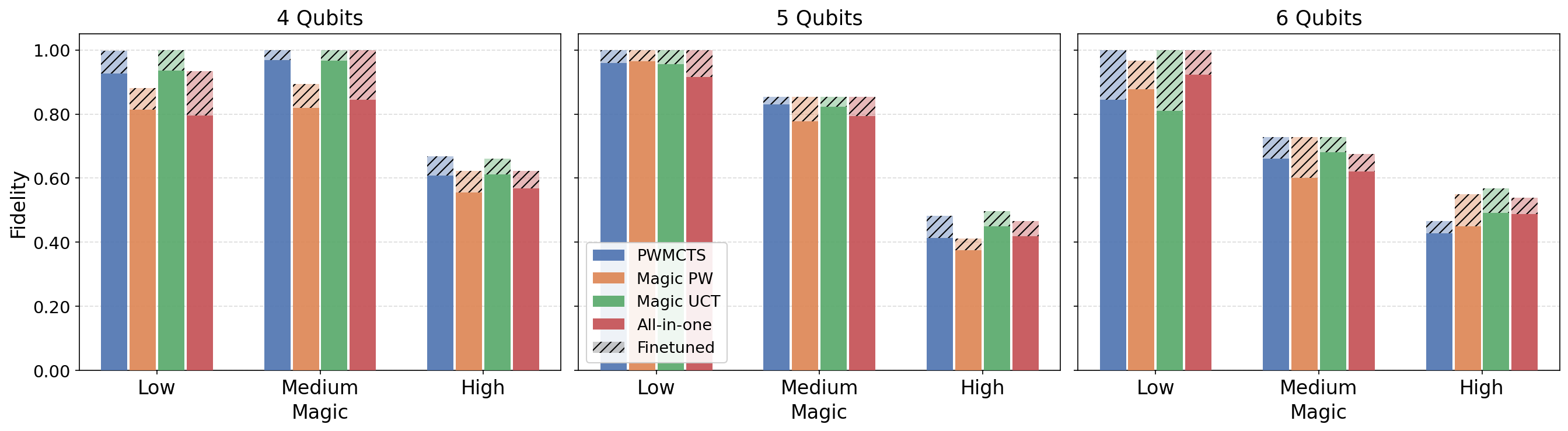}
        \caption{Low-magic PWMCTS variants}
        \label{fig:fidelity_low}
    \end{subfigure}
    
    \caption{Quantum state approximation. The bars report the median values of the fidelity achieved by each PWMCTS variant over 10 independent runs for all the nine target circuits.}
    \label{fig:fidelity_combined}
\end{figure*}

We further evaluate the magic-informed PWMCTS based on the solution quality, measured in terms of fidelity.
Figure~\ref{fig:fidelity_combined} illustrates the numerical results obtained by each PWMCTS variants, including the high-magic class in Figure~\ref{fig:fidelity_high} and low-magic class in Figure~\ref{fig:fidelity_low}. Both figures are divided into three panels, each corresponding to the target problems with fixed number of qubits $n$, and each including the three target problems with increasing magic, corresponding to the target states summarized in Table~\ref{tab:target_circuits}. 
In solid colors there are the result obtained by the gradient-free variants, while in dashed there is the improvement provided by finetuning the circuit parameters using the Adam optimizer.
We observe that the $M_2$ value of the target circuits significantly affects the difficulty of the PQC design problem. In the low-magic regime, the baseline PWMCTS achieves near-perfect performance. However, its performance degrades substantially as the target circuits exhibit higher and more structured magic. This degradation occurs for medium‑magic circuits with at least five qubits and is even more pronounced in the high‑magic regime.

The magic-informed PWMCTS, with the class of high-magic variants, achieves at least the same fidelity as the baseline fo all target states less than the $4$-qubit quantum circuit with low magic, which is the instance with less magic overall. Significant improvements are achieved on target states in the medium- or high-magic regime. For example, on $5$- and $6$-qubit circuits. The ablation study, which allows to analyze the independent contributions from each of the two magic-based biases, highlights that the magic PW is not only stronger in inducing the bias, but it also provides the most significant improvements in terms of solution quality. On the contrary, the PWMCTS variants belonging to the low-magic class do not show any significant improvement with respect to the PWMCTS baseline. The main motivation lies in the particular structure of the PQC provided by these technique, which is analyzed in detail in Section~\ref{structures}.

In the high-magic regime, magic-informed techniques design PQCs with $M_2$ values closer to the target while also achieving improved fidelity compared to the baseline PWMCTS. 

\subsection{Comparison of the circuit structures}\label{structures}
The gradient-free PQC design using PWMCTS is followed by a fine-tuning procedure on the parameters of the circuit using a gradient-based optimizer~\cite{kingma2014adam}. In this context, designing PQCs with high trainability is a key challenge. The experimental results presented in the previous sections suggest that the magic-informed PWMCTS variants provide PQCs with magic values in the desired regime, while also exhibiting more robust convergence to the solution. The improvement in the solution quality becomes more consistent after the finetuning procedure. To analyze this behavior, we compare the trainability of the PQCs designed in the experiments presented in Sections~\ref{subsec:quantum_chemistry} and ~\ref{subsec:quantum_state_approximation}, by computing the Quantum Fisher Information (QFI) matrix. QFI is widely used as a metric of trainability and learnability in PQCs~\cite{cerezo2021variational, zhang2023statistical}, because the eigenvectors associated with its largest eigenvalues identify directions in the parameter space along which the state is most sensitive, and thus most suitable to gradient-based optimization. We employ the trace of the QFI matrix to summarize in a single scalar the measure of the overall trainability of PQCs. Figure~\ref{fig:sre_qfi} and ~\ref{fig:sre_qfi_mol} report the distribution of this measure (shown in dark gray boxplots) for the quantum state approximation and the ground-state energy, respectively.
Across both applications and all problem instances, the high-magic variants of magic-informed PWMCTS consistently produce PQCs with higher $M_2$ values and larger traces of the QFI matrix. Notably, the magic-informed UCT component tends to generate circuits with higher QFI compared to the magic-based Progressive Widening (PW) alone, while their combination achieves the largest overall increases in both $M_2$ and QFI.
However, magic PW exhibits greater robustness in maintaining simultaneously high $M_2$ and QFI values when the GNN operates in out-of-distribution regimes. This behavior is illustrated in Figure~\ref{fig:sre_qfi_mol}. For the H$_2$ molecule—a 4-qubit system where the GNN operates in-distribution—magic PW produces PQCs with higher $M_2$ and QFI trace than the baseline, while magic UCT explores a similar $M_2$ range but achieves a larger increase in QFI. Their combination results in the highest values for both metrics.

Table~\ref{tab:combined_results} summarizes the structure of the PQCs designed the standard PWMCTS and the best magic-informed variant based on the solution quality. It reports the mean and the standard deviation (over the 10 independent runs) of the total number of quantum gates, the number of parameterized gates and the number of the gradient-based Adam optimizer used to finetune the circuits~\cite{kingma2014adam}. The quantum state approximation results have been aggregated in the number of qubits, so each raw of the second part collects the results corresponding to the $4$-, $5$- and $6$-qubit target circuits. 
Low-magic PWMCTS variants consistently design PQCs with fewer total gates and a lower ratio of parameterized gates compared to PWMCTS, indicating a preference for simpler ansatz. In contrast, high-magic variants tend to produce deeper circuits with a higher number of parameterized gates (in ratio) than the baseline, reflecting a wider use of non-Clifford operations.
This behavior is intuitive, as single-qubit rotations are a primary source of magic. However, the comparison with PWMCTS suggests that simply increasing the number of rotations is not sufficient to directly increase the magic of the final PQC, because it requires placing these gates in appropriate positions. This is reflected in the controlled increase of total gates observed in the high-magic regime.
The difference between the PQCs designed by the magic-informed variants and the standard PWMCTS can also be identified in the number of optimization steps required during the finetuning procedure. 
Low-magic variants typically require fewer optimization steps due to the fewer parameters, while high-magic variants provide PQC with more complex structure, whose larger parameter space effectively translates into greater improvements in solution quality.

\section{Discussion}\label{discussion}
The role of quantum magic in variational quantum algorithms is attracting increasing attention~\cite{spriggs2025quantum,capecci2025role}. As current approaches often operate below their quantum potential~\cite{cerezo2025does, bermejo2026quantum, kesiku2026quantum}, we highlight the need to account for quantum resources during the PQC design.
The proposed magic-informed PWMCTS technique leverages a GNN model to efficiently predict the stabilizer $2$-Rényi entropy $M_2$ to control the magic levels during this process~\cite{lipardi2026nonstabilizerness}. The numerical results presented in Section~\ref{experiments} show that the two magic-based biases incorporated in the standard PWMCTS successfully contribute to increase the average magic level in the search tree and most importantly in the final PQC designed. 
The performance of magic-informed QAS substantially depends on the quality of the predictions of the magic estimator. 
An ablation study reveals that the magic PW, which relies on the capabilities of the GNN model to rank PQC based on their $M_2$, is more robust than magic UCT, which relies on the quality of the $M_2$ estimations of the GNN model.

The numerical results also show a benefit of the magic-informed PWMCTS in terms of the solution quality, as it achieves a more robust convergence than the standard PWMCTS. 
In the ground-state energy problem, we tackled three molecules: hydrogen H$_2$, water H$_2$O, and lithium hydride LiH.  We note that the GNN model operates in an extrapolation regime on the number of qubits for the water and lithium hydride molecules, as they require $8$ and $10$ qubits, respectively. On the simpler H$_2$, the class of low-magic PWMCTS provides PQC with improved energy. In contrast, on the more complex molecules of H$_2$O and LiH , the PWMCTS variants providing systematically improved solution belong to the high-magic class. This behavior is consistent with commonly used ansatz in variational quantum eigensolvers, such as UCCSD~\cite{lee2018generalized} and ADAPT-VQE~\cite{grimsley2019adaptive}, where the magic of the PQC typically increases during the optimization process. In the quantum state approximation setting, we consider nine benchmark problems generated according to the procedure described in Section~\ref{subsec:testbed}. These instances provide a diverse testbed for evaluating the magic-informed PWMCTS, as they span a range of circuit sizes and target magic levels.
In this setting, the proposed technique achieves improved solution quality when the target state lies in a high-magic regime. This improvement is reflected not only in fidelity but also in the ability to produce circuits whose magic values match more closely those of the target state. This is particularly relevant, as the reward function does not explicitly account for such properties, and high fidelity does not necessarily imply agreement in underlying quantum state properties.

We emphasize that our approach employs magic estimations to induce bias in the search, but the quality of the candidate PQCs is evaluated by reward functions that depend exclusively on the problem. In general, there is no trivial correlation between magic and these reward functions. Therefore, good candidate PQCs may be potentially discarded.

\section{Conclusions}\label{conclusions}
In this paper, we proposed a magic-informed quantum architecture search (QAS) technique to design parametrized quantum circuits (PQCs). Inspired by the AlphaGo approach~\cite{silver2016mastering}, which combines Monte Carlo Tree Search with Deep Learning, we leverage a recently proposed Graph Neural Network (GNN)~\cite{lipardi2026nonstabilizerness} model for predicting the stabilizer $2$-Rényi entropy of PQCs, and integrate it in the Progressive Widening Monte Carlo Tree Search (PWMCTS)~\cite{lipardi2025quantum}. PWMCTS is a problem-agnostic QAS technique that is particularly suitable to support this integration given its gradient-free approach. The resulting magic-informed PWMCTS offers a pathway to more systematically study and exploit quantum resources, which are often underutilized in current approaches~\cite{bermejo2026quantum, kesiku2026quantum}. 

Numerical results on both the structured ground-state energy problem and on the more general quantum state approximation, show that magic-informed PWMCTS can significantly bias the average magic level within the tree and of the final PQC designed. This bias can be induced toward either high- or low-magic circuits, depending on the objective. We note that the proposed technique can effectively bias the search across a range of applications, including settings involving circuits with more qubits than those seen during training (out-of-distribution predictions), demonstrating strong generalization capabilities. While the training data includes circuits with up to $6$ qubits~\cite{lipardi2026nonstabilizerness}, the proposed technique is evaluated on systems with up to $10$ qubits. Additionally, the magic-informed PWMCTS designs more structured PQCs, leading to improved solution quality compared to those designed by the standard PWMCTS.

We identify three main future research directions. The first is the integration of magic estimation into different QAS techniques, such as those based on reinforcement learning and evolutionary algorithms, which may pave the way for resource-aware QAS techniques. The second direction concerns improving the accuracy of GNN-based predictions. The model used in this work is trained on a general dataset of random quantum circuits. However, it could be further fine-tuned on representative set of circuits tailored to the QAS technique employed. The third direction is the extension of magic-informed QAS techniques to real quantum devices. Although the standard PWMCTS exhibits robust behavior under typical noise in NISQ devices~\cite{lipardi2025quantum}, and the employed GNN can be trained to predict the magic measured on noisy quantum devices~\cite{lipardi2026nonstabilizerness}, the practical deployment of magic-informed QAS techniques on NISQ devices requires future investigation.

\section*{Data Availability Statement}
The data and code to reproduce the experiments of this work are publicly available at the GitHub repository: \url{https://github.com/VincenzoLipardi/GNN-MCTS.git}.
\section*{Acknowledgment}
The authors disclose the use of Claude Sonnet 4.6 to assist with the implementation of the experiments in Section~\ref{experiments}. All content was reviewed and edited by the authors, who take full responsibility for the final work.

\bibliography{bibliography}
\bibliographystyle{ieeetr}

\end{document}